\input{aipcheck}

\documentclass[
    ,final            
  ]
  {aipproc}

\layoutstyle{6x9}
\usepackage{sidecap}

\begin{document}

\title{Supernova Polarization \\and the Type IIn Classification}


\author{Jennifer L. Hoffman}{
  address={Department of Astronomy, UC Berkeley, 601 Campbell Hall, Berkeley, CA 94720-3411}
  ,email={jhoffman@astro.berkeley.edu}
}

\begin{abstract}
While the members of the Type IIn category of supernovae are united by the presence 
of strong multicomponent Balmer emission lines in their spectra, they are 
quite heterogeneous with respect to other properties such as Balmer line 
profiles, light curves, strength of radio emission, and intrinsic brightness. 
We are now beginning to see variety among SNe IIn in their polarimetric 
characteristics as well, some but not all of which may be due to
inclination angle effects. The increasing number of known ``hybrid'' SNe with
IIn-like emission lines suggests that circumstellar material may be more common
around all types of SNe than previously thought. Investigation of the correlations 
between spectropolarimetric signatures and other IIn attributes will help us
address the question of classification of ``interacting SNe'' and the possibility 
of distinguishing different groups within the diverse IIn subclass.
\end{abstract}

\maketitle


Type IIn (``narrow-line'') supernovae (SNe IIn) 
are Type II events whose primary distinction is the existence 
of strong narrow hydrogen Balmer emission lines in the spectra 
\citep{Schlegel1990-,Filippenko1997-}. These lines indicate the presence of 
circumstellar material ejected by the progenitor star and excited by 
the UV and X-ray photons from the supernova explosion. The IIn subclass 
includes 2--5\% of all Type II supernovae \cite{Cappellaro1997-}.

Turatto \cite{Turatto2003-} categorized the SNe IIn as a special class of 
core-collapse supernovae related to the SNe Ib/c and hypernovae. 
However, in recent years several new discoveries have blurred the 
boundaries between the IIn category and other types of supernovae.  
It should be noted that since the label of IIn is assigned based on 
spectral characteristics, while that of II-L or II-P is based on features 
of the light curve, these three categories are not mutually exclusive, 
and some objects may be given more than one classification in the literature. 
But observations of ``hybrid'' supernovae such as SN 2002ic \citep{Deng2004-,Hamuy2003-} 
and SN 2005gj \cite{Aldering2006-}, which showed IIn-like H$\alpha$ lines superposed 
on type Ia-like spectra, and of ``chameleon'' supernovae such as SN 2001em 
\citep{Soderberg2004-,Chugai2006-}, which evolved from a Type Ic to a Type IIn over the span of a 
few years, suggest that not only SNe II but potentially all 
supernovae may show signatures of interaction with circumstellar material. 
Consequently, Turatto's revised classification scheme 
\cite{Turatto2007-} includes a broad category of ``interacting SNe'' 
that spans all SN types and includes the SNe IIn.

The IIn category has always been heterogeneous, as noted by Filippenko \cite{Filippenko1997-}, 
whose Figure 14 
presented a non-coeval collection of SN IIn spectra. Figure 1 shows that even when 
compared at similar ages, SNe IIn have quite diverse spectral characteristics. 
The primary feature of SNe IIn spectra, the strong H$\alpha$ emission line, also varies 
substantially in strength and profile between objects of comparable age (Figure 2) 
and even for a given object over time. In particular, the very narrow (FWHM < 
200 km/s) component of H$\alpha$ does not necessarily exist at all times in the 
evolution of a SN IIn. In SN 1997eg (Figure 3; \cite{Hoffman2007b-}), this 
narrow H$\alpha$ 
component disappeared after 100 days post-discovery, was replaced by a small 
symmetric absorption feature, and then reappeared around day 400. 
The H$\alpha$ line in SN 1998S (\cite{Leonard2000-}; 
their Figure 7) also lost its narrow emission component early on, but was very 
asymmetric and changed much more dramatically over time.

\begin{SCfigure}[0.8]
    \includegraphics[width=3.35in]{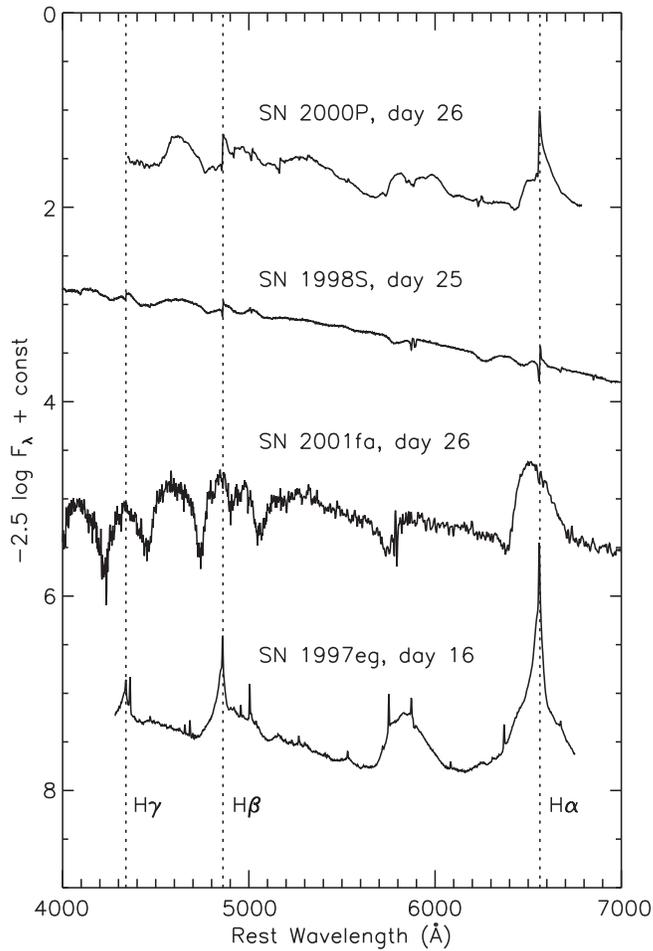}
    \caption{Deredshifted optical spectra of four Type IIn supernovae at similar 
epochs, showing the diversity of spectral characteristics within the subtype. All 
dates are post-discovery. Detailed analyses of SN 1997eg and SN 2000P have been 
conducted by \cite{Hoffman2007b-} and \cite{Leonard2000-}, respectively; the 
other data are presented courtesy of A. Filippenko (priv. comm.). For the 
unpublished spectra, redshift corrections are based on information in the Asiago 
Supernova Catalog \cite{Barbon1999-}. Hydrogen Balmer lines are labeled.} 
\end{SCfigure}

\begin{figure}
    \includegraphics[width=3in,angle=90]{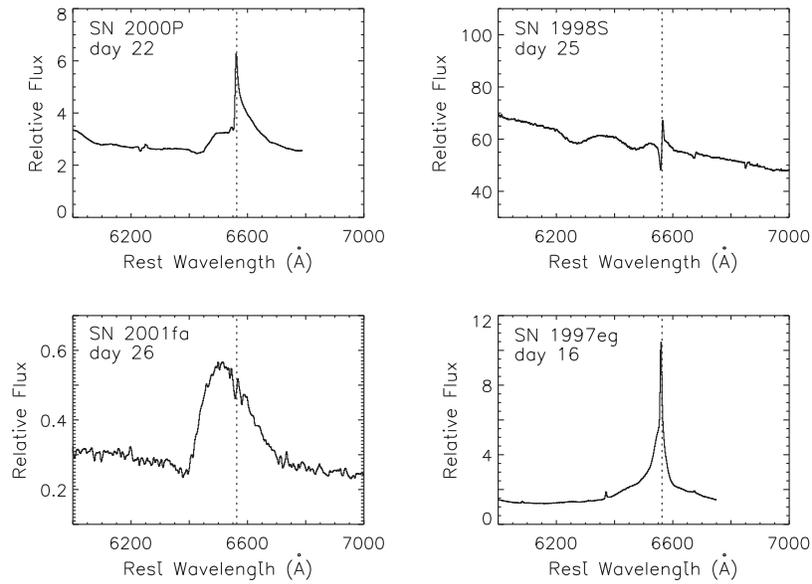}
    \caption{As in Fig. 1, but for the H$\alpha$ region of the spectrum.}
\end{figure}

\begin{figure}
    \includegraphics[width=3in]{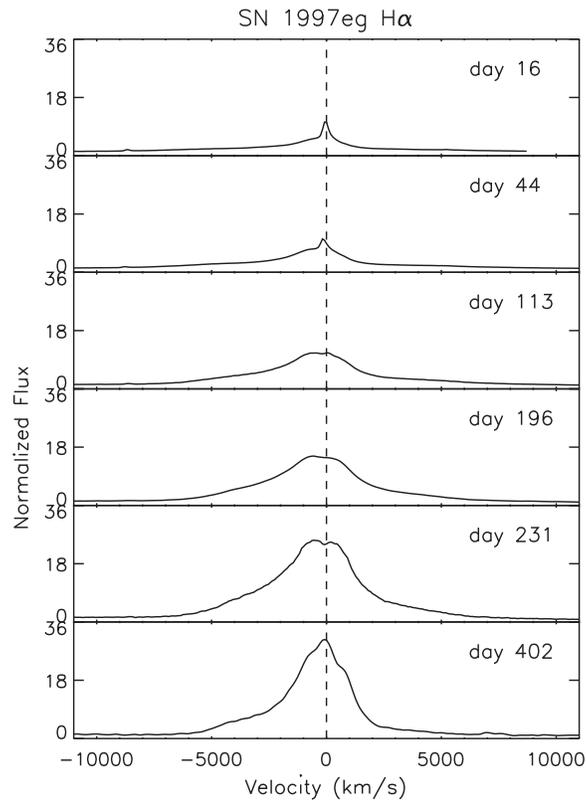}
    \caption{Time evolution of the multi-component H$\alpha$ profile of SN 1997eg
\cite{Hoffman2007b-}. Spectra are normalized to day 16 at $v=-10,000$ km s$^{-1}$.}
\end{figure}

Because of interaction with circumstellar material, the light curves of SNe IIn often
decline quite slowly in comparison with other core-collapse objects, but not all 
members of the subclass show this slow decline \cite{Filippenko1997-}. 
The canonical 
IIn SN 1988Z decreased in brightness by only 5 magnitudes in 1000 days
 \cite{Turatto1993-}, but 
others decline more quickly (e.g., SN 1998S; \cite{Fassia2000-}). SN 1994W faded by 
$\sim$6 magnitudes in {\it B} in only 100 days \cite{Cumming1997-}.

Circumstellar interaction can also cause some SNe IIn to become strong 
radio and X-ray emitters; SN 1988Z and SN 1986J are among the brightest radio supernovae 
ever observed \cite{Chevalier2006-}. Nearly all the SNe IIn detected in X-rays have 
also been strong radio sources \citep{van1996-,Pooley2002-}. However, not all 
SNe 
IIn become radio-loud \cite{van1996-}, and there is considerable heterogeneity even among the 
``radio-quiet'' subset \cite{Filippenko1997-}. The overlap between the IIn, II-P, and II-L categories 
makes it difficult to identify trends in radio brightness with core-collapse subtype.

Finally, SNe IIn show considerable variety in their spectropolarimetric characteristics
(Figure 4). Supernovae of all types are known to be polarized due to intrinsic 
asphericity of the ejecta. The circumstellar material surrounding SNe IIn 
can produce its own polarization signature in addition to that arising from the 
ejecta. Some of the polarization variations between SNe IIn are likely due to 
inclination; that is, they may have similar geometrical distributions of 
circumstellar material, but have different polarization signatures due to different 
viewing angles. To study this effect, I am constructing a grid of Monte Carlo radiative 
transfer models of the H$\alpha$ line polarization produced by circumstellar 
matter distributions with various geometries and seen at various viewing angles 
\cite{Hoffman2007a-}. Preliminary results suggest that differences in viewing angle
may account for some of the polarimetric variety in SNe IIn.
However, such geometric effects cannot explain all 
of the IIn diversity; for example, the transformation of SN 2001em from a Ic into a IIn 
\citep{Soderberg2004-,Chugai2006-} is very unlikely to be due to a sudden change in inclination!

\begin{figure}
    \includegraphics[width=3.7in,angle=90]{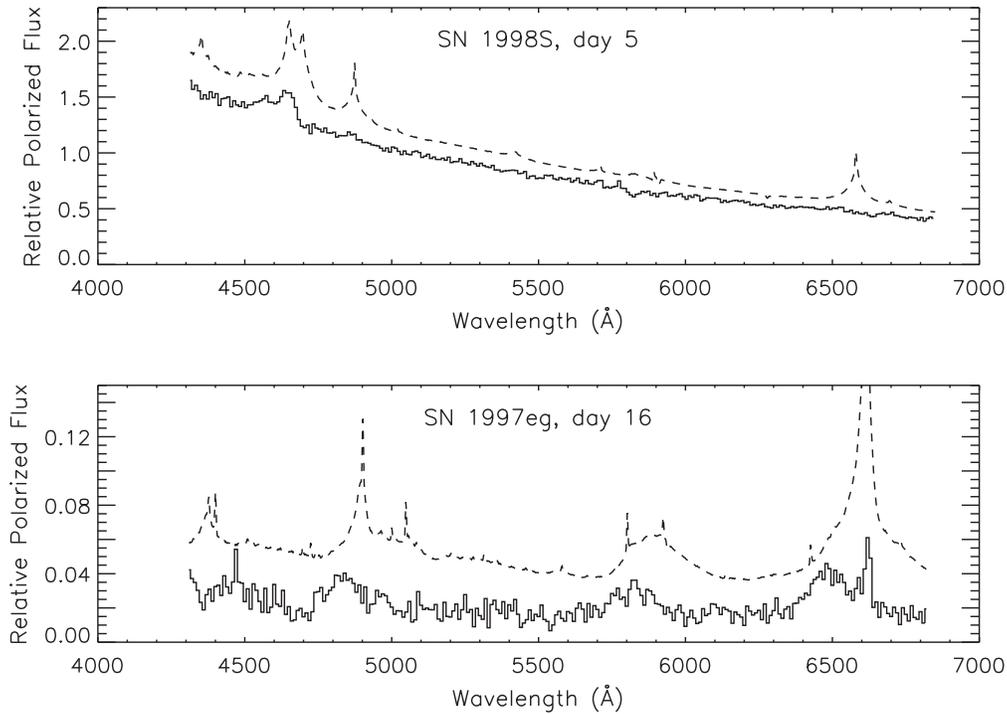}
    \caption{Polarized flux spectra (percent polarization multiplied by flux) for SN 1998S
\cite{Leonard2000-} and SN 1997eg \cite{Hoffman2007b-}, binned to 10 Angstroms in each case.  
Dashed lines show total flux 
spectra for comparison, scaled by 0.037 and 0.036, respectively.  Polarization data are 
corrected for estimated interstellar polarization (ISP) effects; these estimates are 
somewhat uncertain in each case, but a simple correction for ISP could not account for 
all the differences between these two polarized flux spectra.}
\end{figure}

These examples show that the IIn classification is currently something of a ``catchall'' for a 
number of related but distinct objects, all with close connections to Turatto's 
\cite{Turatto2007-} 
new category of ``interacting SNe''. Now that circumstellar interaction has been recognized 
to occur for a broad range of supernova types, and now that high-quality time-dependent 
data are more readily available, we are in a good position to reconsider the 
categorization of SNe IIn and other interacting supernovae. An initiative underway 
within the Berkeley supernova group will quantify the spectral and spectropolarimetric 
characteristics of SNe IIn and search for correlations between these features and other 
properties such as light curve shape and radio/X-ray behavior. If found, such 
correlations will illuminate the relationships between diverse members of the IIn 
category and perhaps ultimately argue for the category's subdivision.

In the future, full understanding and correct classification of these important objects 
will depend critically on obtaining a broad range of observational information, 
including multiwavelength and polarimetric data, with as much time coverage as possible. 
In addition, supernova researchers should expand their collaborations with the evolved star 
community. If the circumstellar media around SNe IIn and related objects are indeed created by 
winds and outflows from massive stars, we can use nearby evolved star nebulae as analogous
cases to the circumstellar envelopes of supernovae. Such an effort could greatly improve our 
understanding of supernova progenitors and help shed light on the physical causes of the 
diversity we observe among interacting supernovae.


\begin{theacknowledgments}
This research was funded by an NSF Astronomy \& Astrophysics Postdoctoral
Fellowship, AST-0302123; by NSF grant AST-0607485 to A. V. Filippenko; and by 
the National Energy Research Scientific Computing
Center, US DOE Contract \#DE-AC03-76SF00098. I thank 
Alex Filippenko at UC Berkeley and Peter Nugent at the Lawrence Berkeley 
Laboratory for their support and invaluable contributions.
\end{theacknowledgments}



\bibliographystyle{aipproc}   

\bibliography{hoffman}

\IfFileExists{\jobname.bbl}{}
 {\typeout{}
  \typeout{******************************************}
  \typeout{** Please run "bibtex \jobname" to optain}
  \typeout{** the bibliography and then re-run LaTeX}
  \typeout{** twice to fix the references!}
  \typeout{******************************************}
  \typeout{}
 }

\end{document}